\documentclass{aastex61}

\shorttitle{3D structure of EUV wave}
\shortauthors{Podladchikova et al.}

\begin{document}

\title{3D reconstructions of EUV wave front heights and their influence on wave kinematics}

\correspondingauthor{Tatiana Podladchikova}
\email{t.podladchikova@skoltech.ru}

\author{Tatiana Podladchikova}
\affiliation{Skolkovo Institute of Science and Technology,
Skolkovo Innovation Center, Building 3, 
Moscow 143026, Russia}

\author{Astrid M. Veronig}
\affiliation{Institute of Physics \& Kanzelh{\"o}he Observatory for Solar and Environmental Research, University of Graz,
Universit{\"a}tsplatz 5,
8010 Graz, Austria}

\author{Karin Dissauer}
\affiliation{Institute of Physics \& Kanzelh{\"o}he Observatory for Solar and Environmental Research, University of Graz,
Universit{\"a}tsplatz 5,
8010 Graz, Austria}

\author{Manuela Temmer}
\affiliation{Institute of Physics \& Kanzelh{\"o}he Observatory for Solar and Environmental Research, University of Graz,
Universit{\"a}tsplatz 5,
8010 Graz, Austria}

\author{Olena Podladchikova}
\affiliation{Solar-Terrestrial Centre of Excellence, Royal Observatory of Belgium}

\begin{abstract}
EUV waves are large-scale disturbances in the solar corona initiated by coronal mass ejections. However, solar EUV images show only the wave fronts projections along the line-of-sight of the spacecraft. We perform 3D reconstructions of EUV wave front heights using multi-point observations from STEREO-A and STEREO-B, and study their evolution to properly estimate the EUV wave kinematics. We develop two different methods to solve the matching problem of the EUV wave crest on pairs of STEREO-A/-B images by combining epipolar geometry with the investigation of perturbation profiles. The proposed approaches are applicable at the early and maximum stage of the event when STEREO-A/-B see different facets of the EUV wave, but also at the later stage when the wave front becomes diffusive and faint. The techniques developed are demonstrated on two events observed at different separation of the STEREO spacecraft (42$^\circ$ and 91$^\circ$). For the 7 December 2007 event, we find that the emission of the EUV wave front mainly comes from a height range up to 90--104~Mm, decreasing later to 7--35~Mm. Including the varying height of the EUV wave front allows us to correct the wave kinematics for the projection effects, resulting in velocities in the range 217--266 km/s. For the 13 February 2009 event, the wave front height doubled from 54 to 93~Mm over 10 min, and the velocity derived is 205--208 km/s. In the two events under study, the corrected speeds differ by up to 25\% from the uncorrected ones, depending on the wave front height evolution.
\end{abstract}

\keywords{Sun: activity --- Sun: corona --- Sun: UV radiation --- waves}

\section{Introduction}
EUV waves are large-scale propagating disturbances observed in extreme ultraviolet (EUV) filtergrams in association with coronal mass ejections (CMEs). After their discovery by the SOHO/EIT instrument \citep{moses97,thompson98}, their characteristics and physics has been intensively debated in the community. In the meantime, there seems to be a consensus reached that EUV waves are large-amplitude fast-mode MHD waves or shocks \citep[cf.\ the recent reviews by][]{patsourakos2012,liu14,warmuth15,long17}.
The disturbance is most likely initiated by the impulsive lateral expansion of the CME flanks, which temporarily act as a piston-type driver \citep[e.g.,][]{Vrsnak2008}. Using the high cadence data from the EUV imagers onboard the STEREO and SDO spacecraft,
it was shown that the EUV wave compression front forms as the CME flanks impulsively expand and then, at a later stage propagates as stand-alone front indicative of a freely propagating blast wave \citep{veronig2008,veronig2010,Kienreich2009,Patsourakos2009b,cheng12}.

The Solar Terrestrial Relations Observatory \citep[STEREO;][]{kaiser08} has been launched with the objective of obtaining stereoscopic observations of the Sun from a near-Earth orbit. The two spacecraft, STEREO-A(head) and STEREO-B(ehind) have slightly different orbits, and thus their angular separation increases by approximately 45$^\circ$ per year. Using stereoscopic data, geometric relations between points in 3D space and their projections onto 2D images can be derived. Data from the STEREO SECCHI suite \citep{Howard2008} provide us with observations of EUV waves from different lines-of-sights \citep[e.g.,][]{Zhukov2009}, giving insight into their 3D geometry and  wave front heights \citep[][]{Kienreich2009,Patsourakos2009b,Temmer11,Delannee2014}
as well as their global propagation characteristics \citep[e.g.,][]{olmedo12}. 

The basic stereoscopy principles for STEREO have been presented by \citet{Inhester2006}. This paper focuses on the problem of the identification and matching of the objects, which is a major challenge in the reconstruction of the STEREO images \citep[see also the review by][]{wiegelmann2009}. As it is shown in \citet{Inhester2006}, the solutions to the matching problem are often strongly tailored to the specific objects observed. A large number of studies has been performed using multi-point STEREO data for stereoscopic reconstruction of objects with well defined structures and/or boundaries such as coronal loops, active regions, polar plumes, jets, prominences or CMEs \citep[e.g.,][]{Inhester2006, Aschwanden2009, Feng2009, Liewer2009, Mierla2009, Temmer2009, Paraschiv2010, Mierla2010, dePatoul2013, Chifu2017}.

Although STEREO data has been extensively used in the analysis of EUV waves 
\citep[e.g.][]{Long2008,veronig2008,veronig2010,veronig2011,Attrill2009,Cohen2009,Gopalswamy2009,Kienreich2009,Ma2009,Patsourakos2009b,Warmuth11,Delannee2014,muhr2014},
the majority of the studies has been carried out \textit{without} exploiting the stereoscopic reconstruction possibility, due to their diffusive nature and the associated difficulties of matching structures in the STEREO image pairs.
The evolution, structure and kinematics of EUV waves have important implications for the understanding of their physical nature. Wave velocities are generally derived by measuring at each time step the position of peak intensity (or outer border) of the wave front propagating in selected directions. As noted by \citet{Ma2009}, it is important to take the actual height of the EUV wave into account, as otherwise the kinematical results will be affected. In addition, line-of-sight projection effects of the optically thin coronal emission can lead to misinterpretation of structures, as e.g., EUV wave fronts versus CME body \citep{Delannee2014}. 

So far, only a few studies on the stereoscopic 3D structure of EUV waves have been performed.
\citet{Kienreich2009} estimated the EUV wave heights  from STEREO quadrature observations on 13 February 2009, when the wave was observed simultaneously against the solar disk by STEREO-B and on the solar limb by STEREO-A. Applying a matching between the kinematics derived from the two vantage points, they estimated the mean height of the wave front to be in the range 80--100~Mm. Height estimations in the more complicated configuration on 7 December 2007 when the spacecraft separation was 42$^\circ$, and STEREO-A and STEREO-B observed different facets of the wave was reported by \citet{patsourakos09}. They used a triangulation technique and studied the wave front at a time when the observations from both spacecrafts became similar, finding a height of about 90 Mm. 
\cite{Delannee2014} is the only study who performed reconstructions of EUV wave heights also in dynamics, i.e.\ studying also the changes in the height during the wave evolution. They tested three different techniques for 3D reconstructions of the wave front, which resulted in quite different results, and present an extensive discussion of the difficulties involved and the problematics of matching. The EUV wave heights they obtain for the 7 December 2007 event are in the range from 35 to 150 Mm, with the wave front height increasing during the initial phase of its evolution and later decreasing. 

In this paper, we develop a technique to reconstruct the 3D structure of the EUV wave front, based on relating the corresponding points in the 2D projections onto the STEREO-A and -B image pairs that are recorded simultaneously.
The main problem to determine the 3D structure of the EUV wave front is the identification of corresponding features and matching of the wave crest in the STEREO-pair images. The epipolar geometry \citep{Inhester2006} is a very useful tool for the 3D reconstruction.	However, it does not give an automatic solution of the matching problem, i.e. the identification of the wave crest on the different STEREO images, and thus additional methods need to be developed. In this study, we present such an approach that is based on a complementary combination of epipolar geometry of stereo vision and investigation of intensity perturbation profiles.  

We present two different solutions to the matching problem of the EUV wave crest on STEREO-pair images. One solution is suitable for the early stage of event development when STEREO-A and STEREO-B see the different facets of the EUV wave and the associated dimming behind. The second one is applicable at the later stage of the event development, when the wave front has similar appearance from both spacecraft, but has become diffusive and faint. We also discuss the conditions in which the matching problem becomes ill-conditioned and degenerates. In this case even small errors can cause distortions of the solution, resulting in unreliable height estimations.
The application of these techniques are demonstrated for two events that occurred on 7 December 2007 \citep[see][]{patsourakos09,Delannee2014} and 13 February 2009 \citep[see][]{Kienreich2009, Patsourakos2009b}. 

\section{Data and data reduction} \label{Data}
We demonstrate and apply the methods developed for the 3D reconstruction to two EUV wave events using image sequences observed simultaneously by the 195\AA~filters of the EUVI instruments onboard STEREO-A and -B with a cadence of  10 min. On 7 December 2007, the separation of the STEREO spacecraft was 42$^\circ$. On 13 February 2009, the EUV wave was observed with the two STEREO spacecraft in perfect quadrature  (at 91$^\circ$ separation). To enhance the wave structures, we constructed EUVI base difference images for the STEREO-A and -B image sequences, and each of the image series was compensated for solar differential rotation. To establish accurate matching of features on the STEREO-pair images and to perform the 3D reconstructions of the EUV wave fronts, the EUVI data were calibrated with the \texttt{SECCHI\_PREP} routines available within SolarSoft, using the precise orbit information from the spice kernels.

The lowest layers observed with the STEREO EUVI imagers correspond to the bottom of the transition region. Thus, for the 3D reconstructions and calculations on the solar sphere we increased the radius of the Sun provided in the image's \texttt{FITS} header (which is the photospheric radius) by 5 pixels, corresponding to about 6 Mm. 
This value of the solar EUV radius relative to the photosphere was derived empirically, by relating small scale features in the EUV images (i.e.\ features with very low heights) to match in STEREO-A and -B images after coordinate transformation. In the following, we use this EUV radius when we refer to the ``surface'' of the solar sphere. 

\section{Methods and analysis} \label{Methods}
We perform 3D reconstructions of EUV wave front heights using multi-point observations from STEREO-A and STEREO-B, 
and study their evolution to properly estimate the EUV wave kinematics. Section~\ref{Stereo_technique} summarizes
the basic idea behind the stereoscopic data analysis technique used for the 3D reconstructions. 
Section~\ref{Wave_Height} describes what we call ``height'' in the context of EUV waves.
The main problem to determine the 3D structure of the EUV wave is the identification and matching of objects on the STEREO-pair images. Sections~\ref{Height_early}~and~\ref{Height_later} present in detail two different solutions 
to the matching problem of the EUV wave crest on STEREO-pair images in order to perform 3D reconstruction.
The methodology is demonstrated on the basis of the 7 December 2007 event.

One solution is applicable at the early stage of event development when the EUV wave is still strong (high amplitude of plasma compression), and STEREO-A and STEREO-B see the different facets of the EUV wave and the associated dimming behind (Section~\ref{Height_early}). The wave crest is identified visually for the first viewpoint and then the corresponding position of the wave crest for the second viewpoint is reconstructed automatically by analyzing the intensity profiles of the perturbation in combination with the epipolar geometry. The second solution is suitable for the later stage of the event development, when the wave front reveals similar appearance from both spacecraft, but has become diffusive and faint (Section~\ref{Height_later}). The wave crest is selected automatically by analyzing the intensity profiles of the perturbation from both satellites STEREO-A and STEREO-B. Finally, the deprojected kinematics of the EUV wave, using the derived height evolution, is presented in Section~\ref{Kinematics}. Section~\ref{Event_2009} presents the results of the proposed approaches applied to 13 February 2009 event.

\subsection{Stereoscopic data analysis technique} \label{Stereo_technique}
\begin{figure}
	\plotone{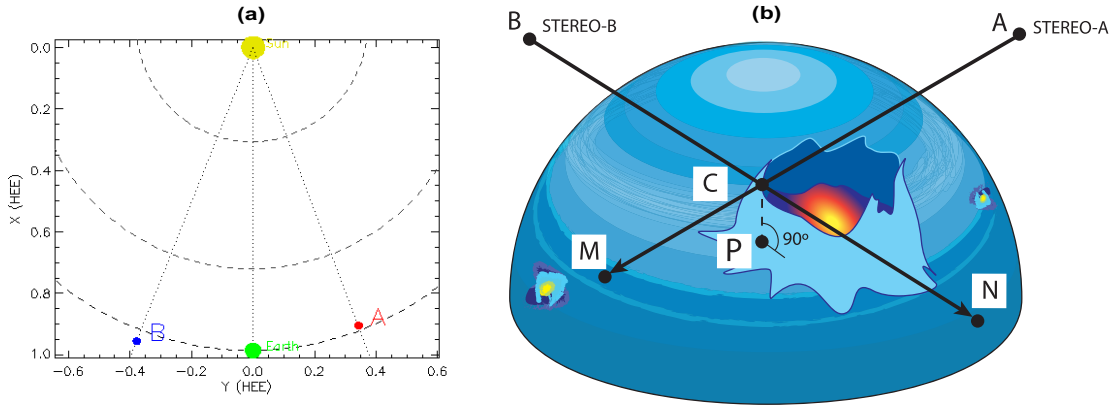}
	\caption{Panel (a): location of STEREO-A and -B on 7 December 2007 with a spacecraft separation of 42$^\circ$.
			Panel (b): schematic illustration of an the EUV wave that is extended in height above the solar surface together with an illustration of triangulation. Point $C$ belongs to the wave crest. Point $P$ is the orthogonal projection of $C$ onto the sphere. Points $M$ and $N$ are the projections of point $C$ onto the sphere along the line-of-sights of STEREO-A ($AC$) and STEREO-B ($BC$).} 
	\label{fig1}
\end{figure}

The height of any feature in the extended solar atmosphere can be reconstructed using the tie-pointing and triangulation technique based on epipolar geometry for STEREO observations as presented in \citet{Inhester2006}.
The twin STEREO-A and -B spacecraft can be considered as the two view points of two observers and provide the opportunity to use stereoscopy for 3D reconstruction for any selected point on or above the Sun’s surface. This technique relies on the identification of identical features in STEREO-A and -B, which is difficult due to very different projections
and line-of-sight integration of the optically thin emission, and is based on epipolar geometry to reduce a two-dimensional problem to a one-dimensional problem. These different viewpoints and projections are illustrated in Figure~\ref{fig1} for  the case of the event on 7 December 2007. 

Figure~\ref{fig1} schematically illustrates the locations of the STEREO-A and STEREO-B spacecraft and an EUV wave with its extent in height above the solar surface together with the triangulation scheme.
The inner side of the eastern wave front (left) and the eastern part of the associated coronal dimming is observed from point $A$. However, this area remains invisible from point $B$. The highest points of the EUV wave crest (as, for example, point $C$) can be viewed by both spacecraft. The line-of-sight from point $B$ to point $C$ is crossed with a sphere (i.e., the solar surface) in point $N$, and the continuation of line $AC$ is crossed with a sphere in point $M$. The point $M$ is a projection of the point $C$ onto the sphere along the line $AC$, and the point $N$ is a projection of point $C$ onto the sphere along a line through $BC$. If the point $C$ can be identified in both images of a simultaneous pair, its location in three dimensions can be determined by computing the intersection of the two lines $AM$ and $BN$. 

The line $BN$ is seen by STEREO-B as a point because it coincides with the line-of-sight from STEREO-B, whereas STEREO-A sees it as an epipolar line in its image plane. Symmetrically, the line $AC$ is seen as a point by STEREO-A  and as an epipolar line by STEREO-B. Therefore the points $C$, $A$ and $B$ form the epipolar plane.
 
The height of the EUV wave front is defined by the length of the segment from the connecting line $CP$, where $P$ is an orthogonal projection of the point $C$ onto the solar surface (a point of intersection with the sphere of the straight line connecting the point $C$ with the sphere center). The change of distance from point $P$ to the eruptive center in time determines the velocity of the EUV wave on the sphere. The distance between the points $M$ and $N$, and the distance between the eruptive center and the wave crest increase with the growth of the height of point $C$ above the surface. If point $C$ is located at a certain height above the surface, then point $M$ is located on the left from point $N$. If point $C$ is located directly on the surface, then points $M$ and $N$ coincide. If the point $C$ is located below the surface, then point $M$ is located on the right from point $N$.

If points $M$ and $N$ are chosen precisely, the straight lines $BN$ and $AM$ in space intersect at point $C$. Errors in the choice of the lines $BC$ and $AC$ lead to definition errors of the coordinates $N$ and $M$. In this case, the straight lines $BN$ and $AM$ may not intersect, thus leading to a failure in the estimation of the coordinates of the point $C$ and its height above the surface.

However, the main source of error that remains is the error in identification of the EUV wave crest in STEREO pair images. These errors depend on the spacecraft separation angle \citep{Liewer2009} as well as on inaccuracy of manual choice of points $M$ and $N$ as usually the 3D-reconstruction results are obtained by manually locating and selecting with a cursor the same feature in both images \citep[e.g.][]{Gosain2009,Delannee2014,Liewer2009,patsourakos09}. As the event progresses, the observations become increasingly noisy and the EUV wave front does no longer have clear boundaries but becomes too diffusive and faint which makes it difficult to reliably detect with the cursor the area of the EUV wave crest. Thus, the estimates obtained this way can be disputable. To reduce these errors and to automatically and reliably define EUV wave crest, we investigate intensity profiles of the perturbation using the ring analysis method of the EUV wave front described in \citet{Podladchikova2005}.

\subsection{Definition of EUV wave height} \label{Wave_Height}
As the plasma is optically thin in the wavelengths we observe, the signal we measure is the radiation emitted by each layer integrated along the line-of-sight (LOS), which makes the identification and matching of corresponding points in the images from the different spacecraft a complicated task. As also discussed in \citet{Delannee2014}, the appearance of the wave front observed from a particular viewpoint thus depends on its geometry and the observing direction. We note that 3D reconstructions in the case of optically thin emission is a nontrivial issue. Actually that holds for any type of analysis of EUV images and EUV waves and any single spacecraft measurement is affected by it. In our study, we make use of the dual spacecraft view from the two STEREO satellites, which increases our information on the object under study due to \textit{two} LOS measurements, with the aim to find the matching points from the observations from both spacecraft. If the emission would be optically thick, then the 3D reconstruction of the height of the object would be a comparatively easy task. But the point of the current study is to find ways how we can perform 3D reconstructions and find matching solutions also in the case of the optically thin emission.

In general, the highest amplitudes of the EUV wave perturbation profiles (as observed from the given observing direction) are interpreted to indicate the wave crest, i.e. the location where the amplitude of the plasma compression is highest 
\citep[e.g.][]{Wills-Davey1999, Podladchikova2005, patsourakos09, Muhr2011}.
As discussed above, this is not a straightforward issue, due to the LOS integration of the emission, the (unknown) shape of the wavefront and also because of temperature changes related to compressions at the wave front. The assumptions we make here are that the wavefront is roughly in radial direction to the surface, that the temperature enhancements at the wave front are small and the typical wave front temperature lies well within the major sensitivity of the 195\AA~passband used (which is supported by the temperature analysis of EUV wave fronts in \citet{Vanninathan2015}). 
We then determine the amplitude of the perturbation profiles as function of distance from the source center independently derived from the two
STEREO satellites, which have different LOS integration, and then use these independent dual measurements to determine the height. With this approach we reconstruct the locations of the largest amplitude (i.e. largest compression) of the disturbance, using the emission enhancement we observe from two LOSs. The main emission enhancement at the wave front will in general stem from a certain height range in the corona (which relates to the height of the wave front). The peaks of the perturbation profiles may not necessarily be indicative of the largest height from where EUV wave emission is radiated (though this may coincide). But in any case they signify the locations (distances from the source region) where we observe the highest wave amplitude, which defines the wave crest.

This means that what we call ``height'' in the context of EUV waves, refers to the height up to which most of the emission that we observe is coming from (the same holds for  previous studies on EUV wave heights; \citet{patsourakos09, Kienreich2011, Delannee2014}). There may be fainter parts of the EUV front existing higher up, consistent with the general interpretation that they are 3D fronts extending to the corona. But we basically determine the height as the range from where the EUV wave fronts show most of its emission. Note that this also implies that we determine the height in the corona which is most significantly affected by the propagating wave disturbance.

\subsection{Solution of matching problem of the wave crest on STEREO-pair images at the early stage of the event}\label{Height_early}
In this section we present our approach to estimate the EUV wave front height above the surface at the early stage of event development when the EUV wave is still strong and images of the wave from the two STEREO spacecraft differ essentially in their observations of the EUV wave and the dimming region behind. We present step-by-step the solution to the matching problem of the EUV wave crest on STEREO-pair images and the algorithm of 3D reconstructions that is based on a complementary combination of epipolar geometry and the investigation of intensity perturbation profiles to determine the wave crest.

\begin{figure}
	\epsscale{0.6}
	\plotone{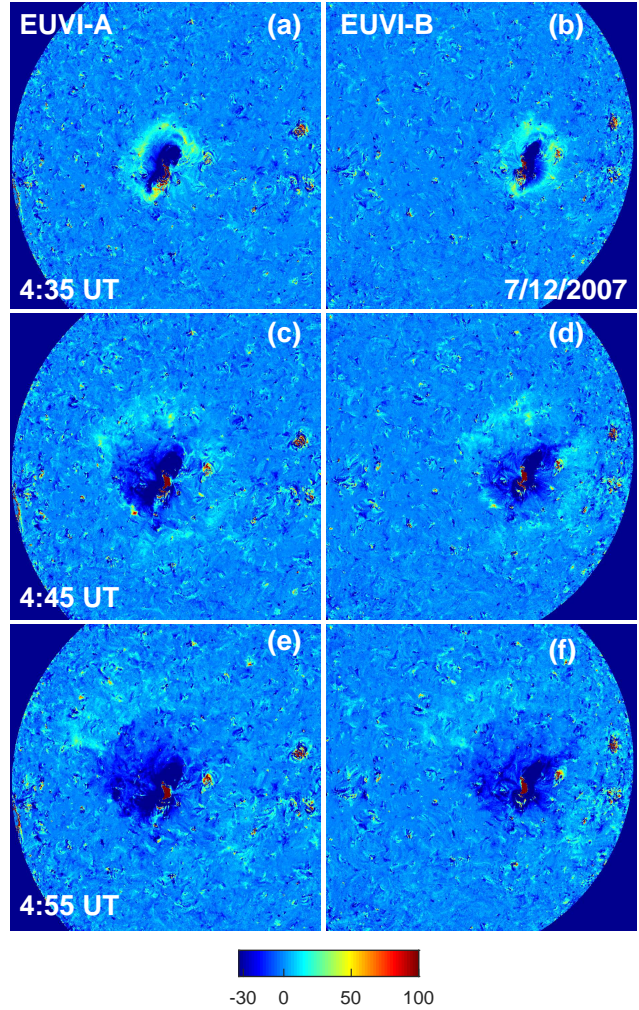}
	\caption{Snapshots of the evolution of the EUV wave of 7 December 2007 from 4:35 to 4:55~UT, as observed in EUVI-A (left) and EUVI-B (right) base-difference images relative to the frame recorded at 4:15~UT. The colors indicate the intensities measured in the 195\AA~filter of the EUVI-A and EUVI-B instruments.}
	\label{fig2}
\end{figure}

Figure~\ref{fig2} shows the dynamics of the EUV wave on 7 December 2007 during 4:35--4:55~UT in base-difference images relative to 4:15~UT.
The left panels show the STEREO-A view, revealing the location of the event in the eastern hemisphere close to the disk center. At that time, STEREO-A was at a distance of 0.967~AU from the Sun and 20.6$^\circ$ West of Earth. The right panels  
show the STEREO-B's view of event (in the western hemisphere), which was at a distance of 1.027 AU and 21.4$^\circ$ East of Earth.

Figure~\ref{fig2}a-b show the STEREO-A and -B views of the early stage of the EUV wave evolution and the associated coronal dimming trailing the wave. This early stage of the EUV wave development at 4:35~UT differs significantly when viewed from the different STEREO vantage points. From the STEREO-A perspective (panel a), the outer boundary of the western segment of the EUV wave front is seen to the right, the inner side of the eastern segment of the wave front to the left, and the dimming to the East. Symmetrically, from the STEREO-B perspective (panel b), the outer front of the eastern segment of the wave front is observed to the left, and the inner side of the western segment of the wave front to the right. The wave crest border is characterized by a sharp gradient in brightness and heterogeneity of texture. If such features, like bright edges, can be correctly identified in the STEREO-pair images, the reconstruction can be performed. The matching problem of the wave crest then consists of correctly identifying these edges in the STEREO-pair images.

The subsequent frames (Figure~\ref{fig2}c-f) show the further development of the event at 4:45 and 4:55~UT. As the EUV wave evolves, its signatures become more similar when viewed from different viewing points exhibiting a quasi-circular character of expansion. However, the wave front becomes more diffuse and is very faint as compared to the early evolutionary stage. The solution of the matching problem of the wave crest on STEREO-pair images in these conditions is presented in the Section~\ref{Height_later}.

\begin{figure}
	\epsscale{0.8} 
	\plotone{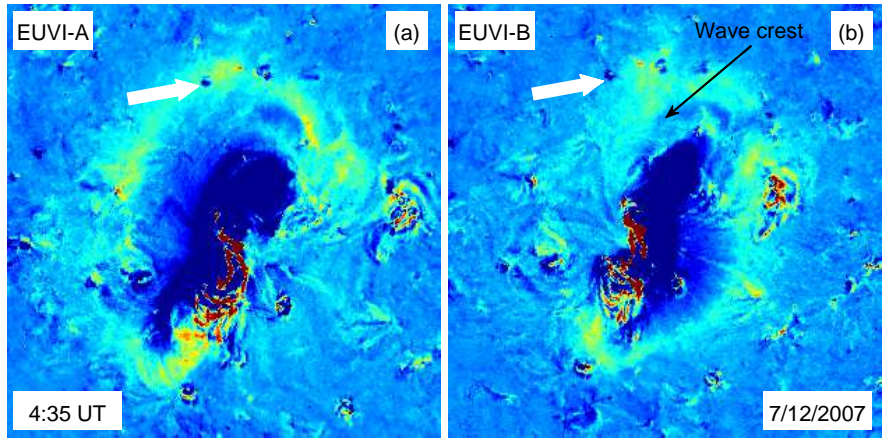}
	\caption{Identification of the EUV wave crest on 7 December  2007, 4:35~UT. The arrow indicates the location of a dark feature identified in STEREO-A (a) and STEREO-B (b).}
	\label{fig3}
\end{figure}

Figure~\ref{fig3}a-b show the eruption at 4:35~UT as viewed from STEREO-A and STEREO-B, respectively, where we identify three small dark features and the left one is marked by a white arrow. These features reveal a specific configuration, and thus they can be uniquely identified and matched in the STEREO-A and STEREO-B image pairs.
As can be seen, the line-of-sight of STEREO-A directed to the left dark feature goes through the brightest part of EUV wave front in the area of wave crest. At the same time STEREO-B sees it \textit{ahead} of the wave front.    
From STEREO-B we observe the whole outer eastern wave front from the indicated feature (which is located on the surface) to the wave crest. A clear boundary of the wave crest is characterized by the sharp decrease of brightness and is indicated by black arrow in the Figure~\ref{fig3}b.
\begin{figure}
\plotone{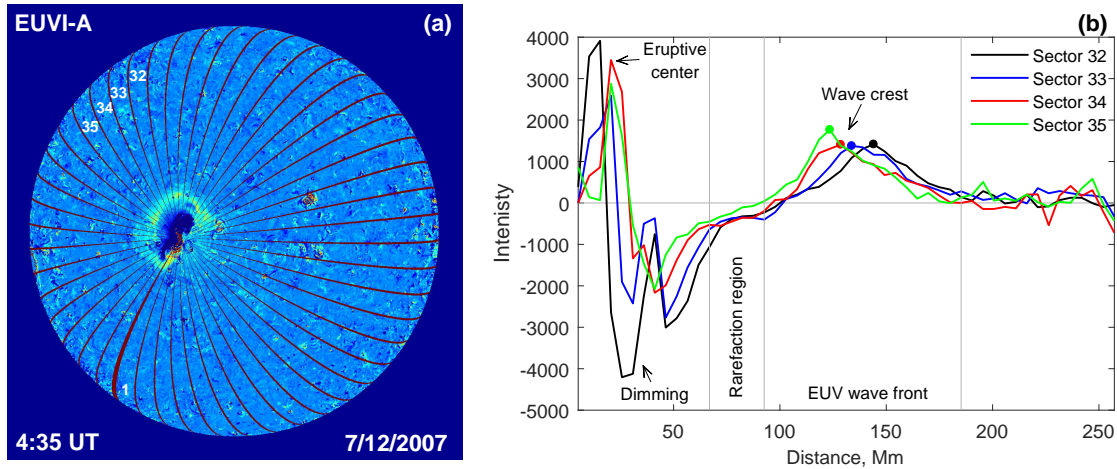}
\caption{Illustration of intensity perturbation profiles used to identify the wave crest.
(a) Event as viewed from STEREO-A at 4:35~UT, showing the division into 48 angular sectors through the eruptive center with a width of $\Delta\phi=7.5$ degrees used to derive intensity perturbation profiles. Sectors 32--35 cover the direction of the EUV waves propagating undisturbed by active regions and coronal holes. (b) Perturbation profiles in sectors 32--35, i.e.\ integral intensity as function of distance from the eruptive center.}
\label{fig4}
\end{figure}

We focus our analysis only on a direction of direct EUV wave propagation with good signal and undisturbed by active regions, coronal holes and other plasma structures. This is important, since for the careful analysis and interpretation of the obtained results in terms of the EUV height the enhanced emission in base difference images needs to stem predominantly from the EUV wave. And not, e.g., from the interaction with other structures passed by the wave, AR loops that are moved (due to oscillations initiated by the EUV wave) or stationary fronts that may be formed at the boundaries to active regions or coronal holes, presenting topological separatrices of the coronal field \citep[see, e.g., review by][]{warmuth15}.  

Figure~\ref{fig4}a shows the event at 4:35~UT as viewed from STEREO-A. To determine the position of the wave crest, we  investigate intensity perturbation profiles using the ring analysis method described in \cite{Podladchikova2005}. We first construct a spherical polar coordinate system with its center on the eruption site. Then the STEREO-A image is divided into 240 rings of equal width of 5145~km around the eruptive center. At this time the EUV wave front is strong in emission and clearly defined, and thus a fine binning using a large number of rings is used which allows us to achieve a better accuracy in the height derivation. As the event evolves, a lower resolution is used for integration of the intensity of the diffusive and faint EUV wave fronts. To select the regions of interest, we additionally define 48 angular sectors with a width of $\Delta\phi=7.5$. The thick brown line in Figure~\ref{fig4} defines the start of our numbering of the sectors, increasing in counterclockwise direction. Sectors 32--35 cover the main regions of the direct wave propagation, i.e. where it propagates without interactions, while the other areas are disturbed by some interactions with small ARs, coronal holes and small-scale magnetic features. In the next step, we calculate for each sector in the base difference images the integrated intensity with the chosen binning of the rings, also called perturbation profiles (as they only reflect the changes with respect to the pre-event base image). The outer border of every ring element is related with the corresponding distance from the eruptive center. As a result, we obtain the projections of the radial intensity profiles onto the surface along the line-of-sight of STEREO-A. 

Figure~\ref{fig4}b shows the dependence of the integral intensity on the distance from the eruptive center in sectors 32--35 as observed by STEREO-A. From the obtained intensity profiles we can define the eruptive center, the dimming regions, and the EUV wave front. 
Close to the eruptive center, we see regions that include segments of maximal brightness due to the associated flare as well as segments of minimal intensity from the coronal dimming, which results from the density depletion due to the  expansion and evacuation of coronal plasma in the wake of the erupting CME
\citep[e.g.,][]{Hudson1996,thompson98,Dissauer2018}. Within the distance range from 67 to 93~Mm (marked by the first two vertical lines in Figure~\ref{fig4}b), the intensity of these regions starts to increase more smoothly with fluctuations around a small negative level. This can be interpreted as the rarefaction region which propagates behind the wave pulse and can be seen as a small, localized dip in the rear section of the intensity profile between the wave pulse and the coronal dimming \citep[see,][]{Muhr2011,Vrsnak2016}.
The intensity of the wave pulse increases sharply from its inner borders at a distance of 93~Mm to a maximum (wave crest) at a distance of 123--144~Mm from the eruptive center. The outer border of the expanding wave pulse is derived at a distance of $\sim$185~Mm as the profile decreases back to the background level. As the background level shows variations around the zero level, an accurate detection of the inner and outer wave boundary is difficult. Thus, by analyzing the intensity perturbation profiles we can determine the projection of EUV wave crest onto the sphere along the line-of-sight of STEREO-A in the considered angular sectors. As it can be seen from Figure~\ref{fig4}b, these EUV wave crest projections marked by colored dots for every angular sector belong to the EUV wave front for the given line-of-of-sight direction of STEREO-A. 

\begin{figure}
	\epsscale{0.8}  
	\plotone{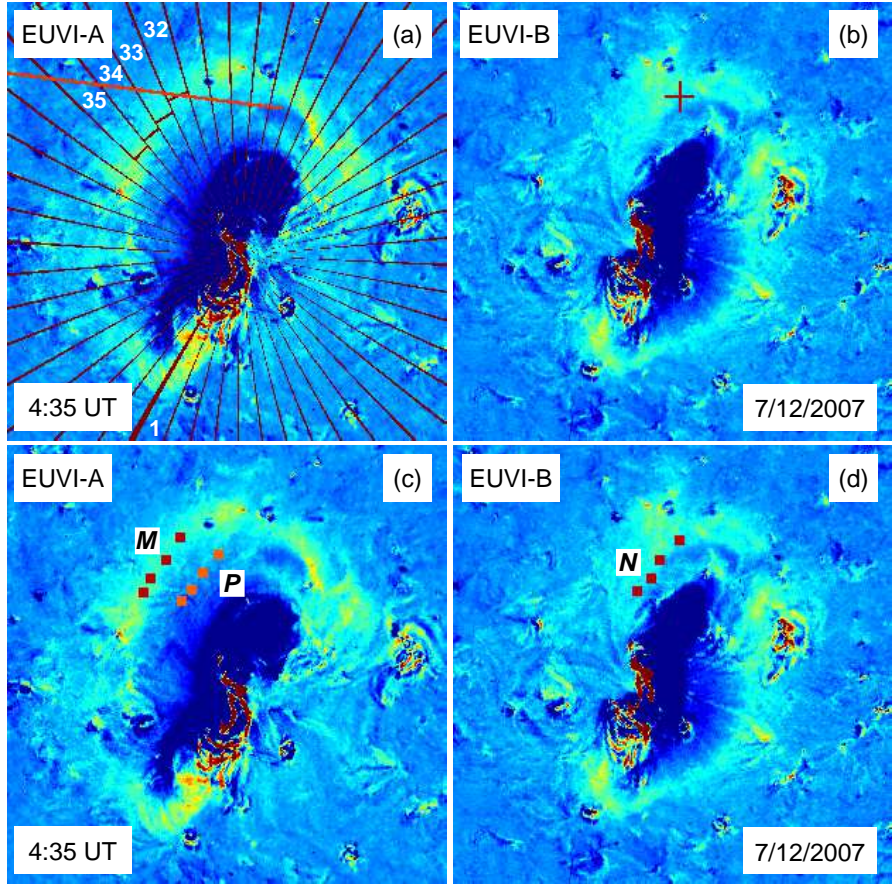}
	\caption{Top panel: Solution of matching problem of the wave crest on STEREO-A (a) and STEREO-B (b) on 7 December 2007, 4:35~UT, when both spacecraft see various facets of the wave. Red cross indicates the chosen point $N$ of the wave crest in STEREO-B (b). The corresponding point $M$ at STEREO-A (a) is found as the intersection of the epipolar line $BN$ (red) as viewed from STEREO-A with the wave crest (brown, sector 32).
	Bottom panel: Height estimation of the EUV wave crest and its orthogonal projections onto the surface for the sectors of direct wave propagation. The selected points $N$ (brown) in the STEREO-B image along the wave crest on 7 December 2007, 4:35~UT are shown at panel (d). The corresponding points $M$ (brown) in the angular sectors 32--35 in the STEREO-A image are indicated in panel (c). The orange points in the STEREO-A image (a) show the orthogonal projections $P$ of the points $C$ onto the surface.}
	\label{fig5}
\end{figure}

In order to perform 3D reconstructions of the EUV wave front,
we first identify the wave crest in STEREO-B image by analyzing the difference in morphology in the image pairs (Figure~\ref{fig2}a-b) and the features in the different image views (Figure~\ref{fig3} and related explanations), and then select a point $N$ of the wave crest, as shown by red cross in Figure~\ref{fig5}b. 
Then, we construct the equation of line $BN$ in space as viewed from STEREO-B:
\begin{equation} \label{Eq_BN}
\frac{X-X_{BB}}{X_{NB}-X_{BB}}=\frac{Y-Y_{BB}}{Y_{NB}-Y_{BB}}=\frac{Z-Z_{BB}}{Z_{NB}-Z_{BB}} \, .
\end{equation}
Here, $(X_{NB},Y_{NB},Z_{NB})$ are the coordinates of the selected point $N$ as viewed from STEREO-B and $(X_{BB},Y_{BB},Z_{BB})$ are the coordinates of point $B$ (STEREO-B) in the RTN coordinate system. The $X$-axis points from Sun center to the respective spacecraft (A or B); $Z$-axis is in the direction of the solar rotational axis, and the $Y$-axis is the cross product of $Z$ and $X$, and lies in the solar equatorial plane (pointing towards the West limb).
The set of points with the coordinates $(X,Y,Z)$, including any sought point $C$ of the wave crest, belong to the equation of line $BN$. Thus, we can consider the set of valid points $X$ and corresponding values $Y$ and $Z$ according to Equation~(\ref{Eq_BN}) that define the ensemble of potential points $C$ of the wave crest as viewed from STEREO-B. 
For convenience the given ensemble of points is denoted by the set of coordinates $(X_{CB},Y_{CB},Z_{CB})$ that belong to line $BN$. We further transfer the whole ensemble of potential points $C$ from STEREO-B to STEREO-A view and derive the ensemble of potential points $C$ with the coordinates $(X_{CA},Y_{CA},Z_{CA})$. This allows us to construct the ensemble of equation of lines $AC$ in space
\begin{equation} \label{Eq_AC}
\frac{X-X_{AA}}{X_{CA}-X_{AA}}=\frac{Y-Y_{AA}}{Y_{CA}-Y_{AA}}=\frac{Z-Z_{AA}}{Z_{CA}-Z_{AA}} \, .
\end{equation}
Here, $(X_{AA},Y_{AA},Z_{AA})$ are the coordinates of point $A$ (STEREO-A); points with the coordinates $(X,Y,Z)$ belong to line $AC$. 
For every potential point $C$ with coordinates $(X_{CA},Y_{CA},Z_{CA})$
the line $AC$ intersects with the sphere at point $M$ described by
\begin{equation} \label{Eq_Sphere}
X^2+Y^2+Z^2=R^2.
\end{equation}
Here, $(X,Y,Z)$ are the unknown coordinates of point $M$ and $R$ is the radius of the sphere. We find the solution of the system of Equations (\ref{Eq_AC},\ref{Eq_Sphere}) for the set of points $(X_{CA},Y_{CA},Z_{CA})$.
The set of solutions represents the ensemble of potential points $M$ with coordinates $(X_{MA},Y_{MA},Z_{MA})$. As a result, the determined ensemble of points $M$ creates the epipolar line $BN$ as viewed from STEREO-A (red line in Figure~\ref{fig5}a). At the same time,
the areas of EUV wave crest projections determined from the analysis of perturbation profiles (Figure~\ref{fig4}b) are indicated by brown lines in Figure~\ref{fig5}a. A point $M$ that corresponds to the selected point $N$ in the STEREO-B image is further found as the intersection point of the epipolar line with the area of the wave crest (brown line, Figure~\ref{fig5}a) for each angular sector 32--35. This means that we graphically found a solution of the system with two equations, one of which characterizes the border of the wave crest and another describing the epipolar line. The combination of epipolar geometry and the investigation of intensity perturbation profiles allows us to reduce the errors of visual choice, and to reliably solve the matching problem of points $M$ and $N$ from the STEREO image pairs.

Figure~\ref{fig5}d shows all the selected points $N$ (brown) along the wave crest in the STEREO-B image, and Figure~\ref{fig5}c gives the corresponding points $M$ (brown) in the STEREO-A image (panel a) on 7 December 2007, 4:35~UT in angular sectors 32--35.
In order to determine the height of the EUV wave crest, we construct two lines $AM$ and $BN$ and find their intersection point $C$ (Figure~\ref{fig1}). The height of the EUV wave is estimated as the distance of point $C$ to the point of intersection with the sphere (line $CP$). We also determine the orthogonal projections $P$ of the point $C$ onto the surface, that are shown by the orange points in Figure~\ref{fig5}c, located in sectors 30--33 in the STEREO-A image, 
and calculate the distance from the point $P$ to the eruptive center. 

Table~\ref{Table1} summarizes the results of the height estimations for all considered points $M$ and $N$ in sectors 32--35 belonging to the wave crest (Height $C-P$) and distances from the orthogonal projections $P$ and points $M$ to the eruptive center (Distance $P-Ec$ and Distance $M-Ec$) at 4:35~UT. We also indicate the number of angular sectors where every point $P$ is located. As can be seen from Table~\ref{Table1}, the height of the EUV wave crest in the chosen sectors varies from 90 to 104~Mm. The points $M$ are located at distances of 119--139~Mm from the eruptive center, while the orthogonal projections $P$ for areas of the direct waves in these sectors are on average 20\% closer to the eruptive center (93--117~Mm). This should be taken into account for the accurate estimations of EUV wave kinematics, as the change of the distance $P-Ec$ in time determines the velocity of the EUV wave on the sphere.
\begin{deluxetable}{c c c c c c}
	\tablecaption{Characteristics of the EUV wave of 7 December 2007, 4:35~UT.
    `Points'' refers to a number of points $N$ and $M$ indicated by brown color in Figure~\ref{fig5}cd. ``Height $C-P$'' gives the height of the EUV wave determined as distance from point $C$ to $P$. ``Distance $P-Ec$'' and ``Distance $M-Ec$'' indicate the distance from the orthogonal projections $P$ and points $M$ to the eruptive center. ``Sector'' gives the number of the angular sector in which the point $P$ is located. 
    \label{Table1}} 
	\tablehead{
		\colhead{Time} &
		\colhead{Points} & 
   	    \colhead{Height $C-P$ (km)} &
		\colhead{Distance $P-Ec$ (km)} &
	    \colhead{Distance $M-Ec$ (km)} &	
		\colhead{Sector} 
	}
	\startdata
	4:35~UT    & 1  & 103 645 & 116 920  & 139 278  & 30  \\
	           & 2  & 94 236  & 105 888  & 128 134  & 31  \\
	           & 3  & 99 329  & 96 871   & 123 305  & 32  \\
	           & 4  & 89 776  & 93 051   & 118 684  & 33  \\
	\enddata
\end{deluxetable}

We stress that it is important that the epipolar line (red) in Figure~\ref{fig5}a intersects the EUV wave crest, as this allows us to find their intersection point. If an epipolar line almost coincides with the tangent to the EUV wave front, the problem has a multitude of solutions, becomes ill-conditioned and degenerates \citep{Rice81}. Therefore, we can not expect to obtain reliable solutions for the northern and southern segments of the wave front.

\subsection{Solution of matching problem of the wave crest on STEREO-pair images at the later stages of event development} \label{Height_later}
In the previous section we analyzed the 3D structure of the EUV wave when the images from STEREO-A and STEREO-B showed different parts of the EUV wave during its early stage of development. Starting from 4:45~UT (Figure~\ref{fig1}c-f), the EUV wave becomes quasi-circular and appears similar in the observations from both spacecraft. On the one hand, this opens additional opportunities to automatically determine the correspondence between separate regions by a criterion of their similarity, which was not possible before. On the other hand, the EUV wave front does no longer have clear boundaries, and becomes too diffuse and faint to be detected using the method presented in Section~\ref{Height_early}. 

\begin{figure}
\epsscale{0.8}
\plotone{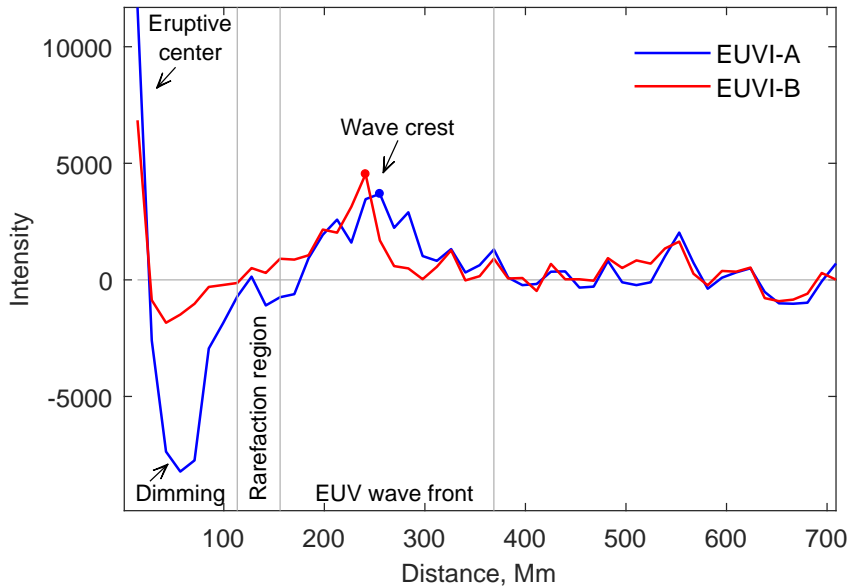}
\caption{Intensity profiles, i.e.\ dependence of the integral intensity on the distance from the eruptive center in sector 30 for STEREO-A (blue) and STEREO-B (red) on December 7, 2007, 4:45~UT}
\label{fig6}
\end{figure}

\begin{figure}
	\epsscale{0.7}
	\plotone{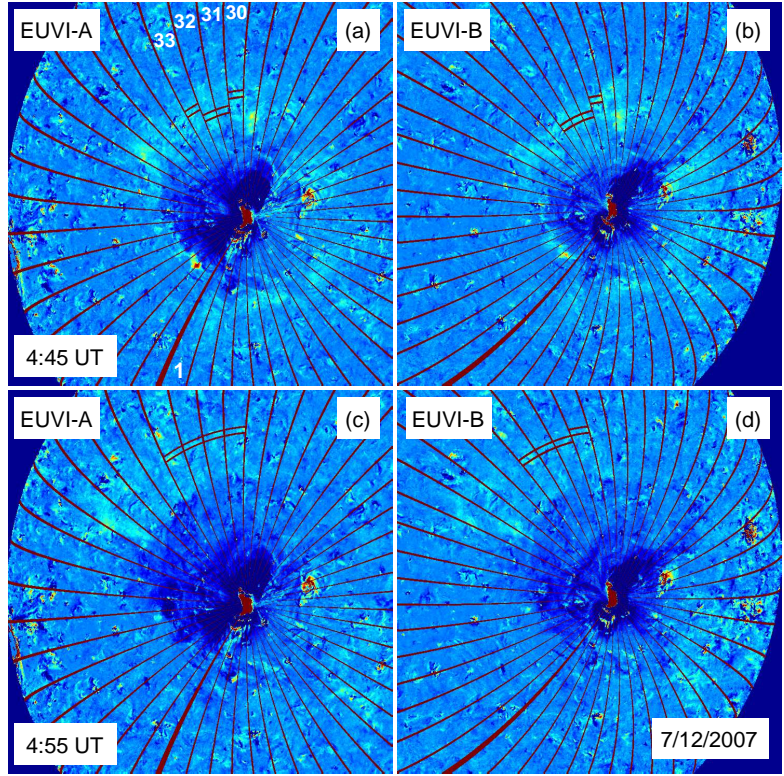}
	\caption{Solution of the matching problem of the wave crest on STEREO-A (left) and STEREO-B (right) on 7 December 2007, 4:45 - 4:55~UT, when the wave front is diffusive and faint. Brown borders show areas of the wave crest in sectors 30--33 determined from the investigation of the intensity perturbation profiles.}
	\label{fig7}
\end{figure}

Therefore, in this section we present an alternative approach to automatically detect the wave crest on STEREO-pair images, and to find the correspondence (matching of points) during these times.  
Analogously to Section~\ref{Height_early}, we investigate intensity profiles, but this time for both spacecraft, STEREO-A and STEREO-B, and with coarser binning of rings to obtain better count statistics. We use 80 rings of equal width of $15 434$~km around the eruptive center, and integrate the intensity in 48 angular sectors along great circles, each spanning a width of $\Delta\phi=7.5$.
From the obtained intensity profiles, we  determine the regions of maximum intensity to define the wave crest.
For illustration, we show in Figure~\ref{fig6}  the intensity profiles from STEREO-A (blue) and STEREO-B (red) for angular sector 30, with the respective location of the wave crest indicated by dots.
Figure~\ref{fig7} shows STEREO-A and -B images at 4:45 UT and 4:55~UT together with the 48 sectors defined and the area of wave crest limited by the ring width (indicated by brown borders for sectors 30--33) as automatically determined by the procedure described above. 
In the following, we focus our analysis on the 3D reconstruction of the EUV wave height on the same sectors 30--33, where the orthogonal projections points $P$ are located at 4:35~UT (Section~\ref{Height_early}), to determine the heights and kinematics of the faint EUV wave during the period 4:45--4:55~UT. 
The STEREO-A and -B pairs of intensity profiles, such as plotted in Fig.~\ref{fig6}, are used to determine the correspondence between small segments of the obtained curves.  Each data point of these profiles characterizes the brightness of the given segment. The $x$-axis gives the segment's location, i.e.\,the location of the points $M$ for the blue line and points $N$ for the red line.

Both spacecraft observe the same features, thus the profiles in Figure~\ref{fig6} have similar forms with the exception of the eastern dimming region that remains partially invisible from STEREO-B. Both STEREO-A and -B profiles show one pronounced local maximum, which we identify as the wave crest (marked by blue and red points). As it can be seen in Figure~\ref{fig6} the projection of wave crest along the line-of-sight of STEREO-A is located at a distance of 255~Mm from the eruptive center, whereas in STEREO-B it is 241~Mm. In this case, both points $M$ and $N$ are located on the left from the eruptive center (Figure~\ref{fig7}), and the point $M$ is located at a greater distance from the eruptive center and thus to the left from the point $N$. According to the scheme in Figure~\ref{fig1}, this means that point $C$ of the wave crest is located at a certain height above the surface and we can use these information to perform 3D reconstructions.

The same procedure is used to establish the correspondence between the segments of the wave crest on both images for the other sectors 31--33 at 4:45~UT and sectors 30--33 at 4:55~UT. Note that the maxima of intensity profiles for STEREO-A and STEREO-B at sector 30 at time 4:55~UT are located at the same distance from the eruptive center, i.e.
points $M$ and $N$ coincide, which indicates that point $C$ is located very close to the surface. This implies that at this late stage of the event, the height range from where we observe significant emission of the EUV wave has decayed to a very small value. Therefore this sector was excluded from further analysis. To determine the corresponding points $M$ and $N$ in each considered segment, we determine the center of the wave crest for STEREO-A and STEREO-B (brown borders, Figure~\ref{fig7}). Similarly we determine the 3D coordinates of point $C$ (the highest point of the  wave crest,  Figure~\ref{fig1}) by solving the system of two equations, one of which is the equation of line $AM$, and another is the equation of line $BN$. The height of the EUV waves is then estimated as a distance from the point $C$ to the point $P$.

\begin{deluxetable}{c c c c c c}
	\tablecaption{Characteristics of EUV wave at 7 December 2007, 4:45--4:55~UT. ``Points'' gives a number of points $N$ and $M$ corresponding to the center of each area of the wave crest for STEREO-A and STEREO-B (brown points, Figure~\ref{fig7}). ``Height $C-P$'' gives the height of EUV wave determined as distance from the point $C$ to the point $P$. ``Distance $P-Ec$'' and  ``Distance $M-Ec$'' indicate the distance from the orthogonal projections $P$ and points $M$ to the eruptive center.  ``Sector'' gives the number of the angular sector where the point $P$ is located. \label{Table3}} 
	\tablehead{
		\colhead{Time} &
		\colhead{Points} & 
		\colhead{Height $C-P$ (km)} &
		\colhead{Distance $P-Ec$ (km)} &
  	    \colhead{Distance $M-Ec$ (km)} &	
		\colhead{Sector} 
	}
	\startdata
	4:45~UT    & 1  & 7 047    & 266 303  & 270 066 & 30  \\
			   & 2  & 8 895    & 236 111  & 240 255 & 31  \\
               & 3  & 12 046   & 234 420  & 239 758 & 32  \\
               & 4  & 35 198   & 252 834  & 270 991 & 33  \\
    \hline
    4:55~UT    & 2  & 10 378   & 386 617  & 394 249 & 31  \\
               & 3  & 12 220   & 385 440  & 394 015 & 32  \\
               & 4  & 13 557   & 383 833  & 394 802 & 33  \\
	\enddata
\end{deluxetable}
Table~\ref{Table3} summarizes the results of the height estimations in the sectors 30--33 at time steps 4:45 and 4:55~UT. As can be seen, the height of the EUV wave crest at this later stage of event development at 4:45--4:55~UT has substantially decreased to 7--35 Mm. The points $M$ are located at a distance of 240--271~Mm at 4:45~UT and 394--395~Mm at 4:55~UT from the eruptive center, and the corresponding orthogonal projections $P$ are close to points $N$ (234--266~Mm from the eruptive center at 4:45~UT, 384--387~Mm at 4:55~UT). Note that the discrepancy in distance to the eruptive center of the orthogonal and the line-of-sight projections  compared to that for 4:35~UT is insignificant because the height of the EUV wave front has dramatically decreased over this time interval. 

We note that under the conditions of faint and diffusive EUV wave fronts, the automatic detection of the wave crest provides a more reliable solution in comparison to a visual selection of points on the wave crest. In particular, as the integral intensity of a small segment along the calculated profiles is less disturbed by localized deviations than individual pixels, and thus it reflects more accurately the same regularities of wave crest location on both images.

\subsection{EUV wave kinematics} \label{Kinematics}
The EUV wave as observed in the STEREO images is a projection along the line-of-sight of the spacecraft.
Thus, the kinematics of the EUV wave crest we observe in single spacecraft images may differ from the actual one. The 3D reconstruction of the height of the EUV wave fronts allows us to correct the wave kinematics for projection. In Sects.~\ref{Height_early}~and~\ref{Height_later}, we determined the orthogonal projections of the wave crest onto the surface and their distances from the eruptive center (Tables~\ref{Table1}~and~\ref{Table3}). From these estimations, we can determine the velocity of the EUV wave crest. The results for the propagation in sectors 30--33 over the interval 4:35--4:55~UT are summarized in Table~\ref{Table4}.
We show both, the velocities derived using the orthogonal projections of the wave crest onto the surface and that on the basis of line-of-sight projection. From the comparison of both, we can estimate the errors in the kinematics derived from the line-of-sight projections.

\begin{deluxetable}{c c c c c}
	\tablecaption{Estimation of EUV wave velocities on 7 December 2007, 4:35 - 4:55~UT.
		``Time'' gives the time period over which the velocity of the EUV wave is estimated.
		``Points'' indicates the number of points $N$ and $M$ (brown points, Figures~\ref{fig5}cd~and~\ref{fig7}). ``Velocity (km/s), Orthogonal projection'' gives the estimation of velocity using the orthogonal projections of the wave crest onto the solar surface. ``Velocity (km/s), Line-of-sight projection'' lists the estimation of velocity on the basis of the line-of-sight projection. 
		\label{Table4} } 
	\tablehead{
		\colhead{Time} &
		\colhead{Points} &
		\colhead{\parbox{4cm}{\centering Velocity (km/s) \\ Orthogonal projection}} &
		\colhead{\parbox{4cm}{\centering Velocity (km/s) \\ Line-of-sight projection}} &
		\colhead{Sector} 
	}
	\startdata
	4:35 - 4:45~UT    & 1  & 249 & 218 & 30\\
				  	  & 2  & 217 & 187 & 31\\
   	                  & 3  & 229 & 194 & 32\\
  	                  & 4  & 266 & 254 & 33\\
	\hline \\[-4ex]		
	4:45 - 4:55~UT    & 2  & 251 & 257 & 31\\ 
	                  & 3  & 252 & 257 & 32\\
                   	  & 4  & 218 & 206 & 33\\ 
	\enddata                 
\end{deluxetable}

As shown in Table~\ref{Table4}, the velocity of the wave crest over the time interval 4:35--4:45~UT estimated on the basis of line-of-sight projection (187--254 km/s) is about 5--18\% smaller than that using the orthogonal projection (217--266 km/s). Due to the large height of the EUV wave front of 90--104~Mm at 4:35~UT, the distance between the orthogonal and line-of-sight projections is large. 
Therefore, during the time period 4:35--4:45~UT, the line-of-sight projections of the wave crest covered a smaller distance than the orthogonal ones. This means that the EUV wave velocity determined on the basis of line-of-sight projection is underestimated, and we should use the orthogonal projection to estimate the actual EUV wave velocity on the sphere.
However, after 10 minutes of evolution (at 4:45~UT), the height of the EUV wave crest hast strongly decreased to 7--35~Mm, and thus the line-of-sight projections are very close to the orthogonal ones.  Therefore, the wave crest velocities over the time period 4:45--4:55~UT estimated on the basis of orthogonal and line-of-sight projections differ only slightly ($\lesssim  5\%$). 
This is related to the fact that there is no significant change in the height of the EUV wave crest, and thus the estimations of wave kinematics can be performed on the basis of line-of-sight projections.

\section{Results for the 13 February 2009 event} \label{Event_2009}
In this section we apply our approach presented in the Section~\ref{Methods} to estimate the height and kinematics of the EUV wave of 13 February 2009 (launched from Active Region 11012), which was observed by the two STEREO spacecraft in quadrature.  On 13 February 2009 STEREO-A was at a distance of 0.964 AU from the Sun and 43.7$^\circ$ West of the Earth, and STEREO-B was at a distance of 1.0004 AU and  47.5$^\circ$ East of the Earth. 

\begin{figure}
	\plotone{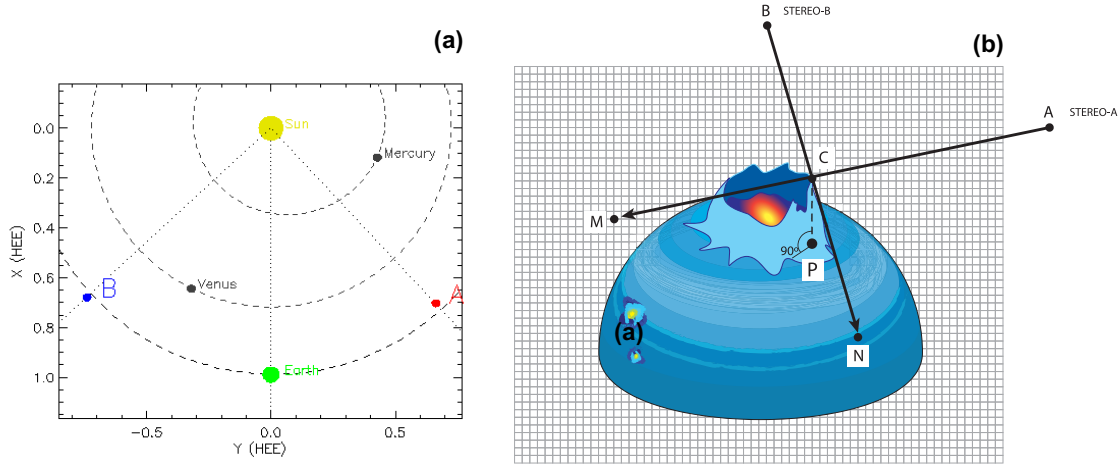}
	\caption{Panel (a): Location of STEREO-A and -B on 13 February 2009, when the EUV wave
		was observed with the two STEREO spacecraft in quadrature.
		Panel (b): Schematic location of STEREO-A and STEREO-B for the 13 February 2009 event 
		together with an illustration of triangulation for the EUV wave.}
	\label{fig8}
\end{figure}

Figure~\ref{fig8} schematically shows the locations of the STEREO-A and -B spacecraft and the EUV wave.
This scheme is similar to that presented in Figure~\ref{fig1} with the only exception that in this case point $M$ may be out of the sphere belonging to the plane $YOZ$, where $O$ is the  center of the Sun, and $Y$ and $Z$ are the axes of the coordinate system of STEREO-A. 

\begin{figure}
\epsscale{0.8}  	
\plotone{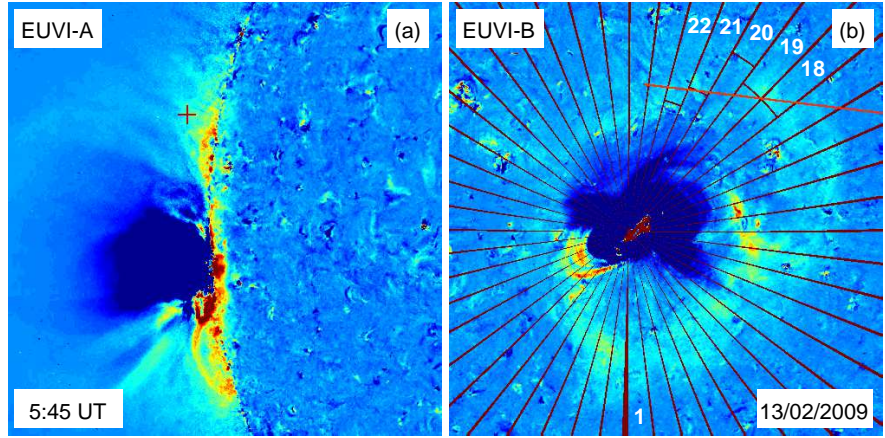}
\caption{Solution of the matching problem of the wave crest on STEREO-A (a) and STEREO-B (b) for 13 February 2009, 5:45~UT. The red cross indicates the chosen point $M$ of the wave crest in STEREO-A (a). The corresponding point $N$ at STEREO-B (b) is found as the intersection of the epipolar line $AM$ (red) as viewed from STEREO-B with the wave crest determined by the maximum in the intensity profiles (brown).}
\label{fig9}
\end{figure}

Figure~\ref{fig9} shows snapshots of the event in base-difference images at 5:45~UT relative to 5:25~UT as well as 
the solution of matching problem of wave crest on STEREO-pair images based on the combination of epipolar geometry and analysis of perturbation profiles. To estimate the height of the EUV wave front and its kinematics,  we first select a point $M$ of the wave crest in the STEREO-A image, as shown by the red cross in Figure~\ref{fig9}a. This point $M$ may belong either to the front or the rear part of the wave crest, depending on their heights.
In the same way as described in Sect.~\ref{Height_early}, we trace the epipolar line $AM$ as viewed from STEREO-B (red line in Figure~\ref{fig9}b) that is now determined by the intersecting line $BN$ (Equation~\ref{Eq_BN}) with the plane $YOZ$ ($X=0$). Then we find the point $N$ in STEREO-B that corresponds to the selected point $M$ in the STEREO-A image as the intersection point of the epipolar line with the area of the wave crest (brown line in Figure~\ref{fig9}b), determined from the investigation of the intensity profiles. As can be seen in Figure~\ref{fig9}b, the constructed epipolar line (red) intersects the areas of the wave crest in sectors 19 and 21. In the following, we concentrate on sector 19, as here the wave exhibits a higher brightness and a propagation undisturbed of  interactions with plasma structures. 

\begin{figure}
\epsscale{0.7}
\plotone{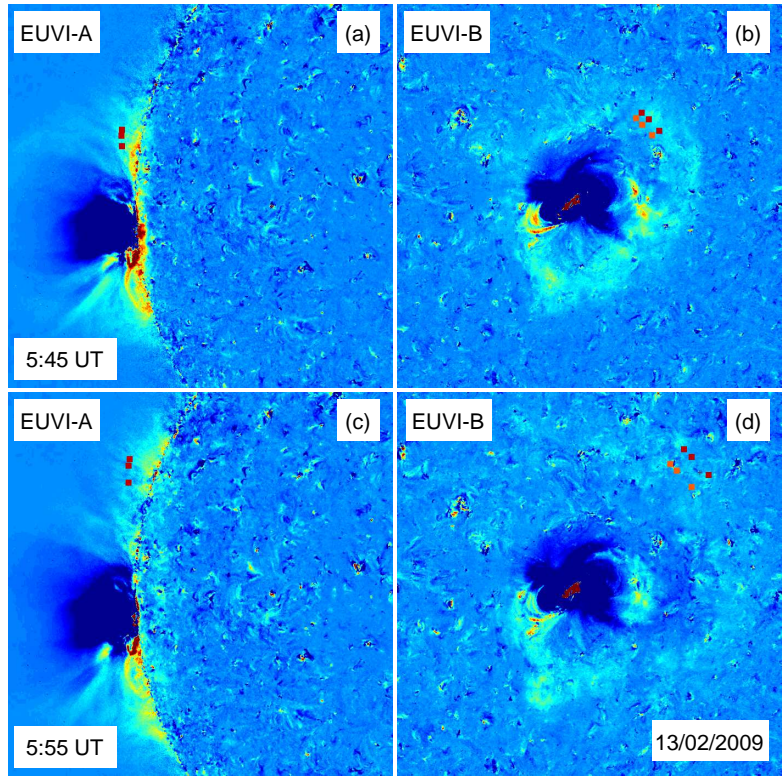}
\caption{Height estimation of the EUV wave crest and its orthogonal projections onto the surface for the event on 13 February 2009, 5:45--5:55~UT. The selected points $M$ (brown) in the STEREO-A image along the wave crest are shown in panels (a) and (c). The corresponding points $N$ (brown) in angular sectors 18--22 in the STEREO-B image are indicated in panels (b) and (d). The orange points in the STEREO-B image (panels (b) and (d)) show the orthogonal projections $P$ of the points $C$ onto the surface.
}
\label{fig10}
\end{figure}

Figure~\ref{fig10} shows all the selected points $M$ (brown) at the STEREO-A image along the wave crest at 5:45~UT (a) and 5:55~UT (c), and the corresponding points $N$ (brown) in the STEREO-B image in angular sectors 18--20. The height of the EUV wave front is estimated as the distance from point $C$ (determined from intersection of lines $AM$ and $BN$) to the point of the intersection with the sphere $P$ (Figure~\ref{fig9}). We also define the orthogonal projections $P$ of the point $C$ onto the surface, that are shown by the orange points in Figure~\ref{fig10}b-d and are located in sectors 18--19 in STEREO-B, and calculate the distance from the point $P$ to the eruptive center. 

\begin{deluxetable}{c c c c c c}
	\tablecaption{Height reconstruction of the EUV wave on 13 February 2009 5:45--5:55~UT. ``Points'' gives a number of points $N$ and $M$ indicated by brown color in Figure~\ref{fig10}. ``Height $C-P$'' gives the height of the EUV wave determined as distance from the point $C$ to the point $P$. ``Distance $P-Ec$'' and ``Distance $M-Ec$'' indicate the distance from the orthogonal projections $P$ and points $N$ to the eruptive center. ``Sector'' gives the number of the angular sector where the point $P$ is located. \label{Table5}} 
	\tablehead{
		\colhead{Time} &
		\colhead{Points} & 
		\colhead{Height $C-P$ (km)} &
		\colhead{Distance $P-Ec$ (km)} &
		\colhead{Distance $N-Ec$ (km)} &	
		\colhead{Sector} 
	}
	\startdata
	5:45~UT    & 1  & 53 894   & 238 207  & 255 400  & 19  \\
			   & 2  & 55 688   & 236 110  & 255 092  & 19  \\
  			   & 3  & 56 442   & 236 627  & 255 189  & 18  \\
	\hline
    5:55~UT    & 1  & 88 213   & 362 713  & 411 900  & 19  \\
			   & 2  & 90 967   & 361 139  & 410 266  & 19  \\
			   & 3  & 93 116   & 359 699  & 410 616  & 18  \\
	\enddata
\end{deluxetable}

Table~\ref{Table5} summarizes the results of the EUV wave height estimation in sectors 18--19. The height of the EUV wave crest almost doubled from 54 to 93~Mm during 5:45 to 5:55~UT. The points $N$ are located at a distance of 255~Mm at 5:45~UT and at the distance of 410--412~Mm at 4:55~UT from the eruptive center, while the corresponding orthogonal projections $P$ are closer to the eruptive center (238--237~Mm at 5:45~UT, 360--363~Mm at 5:55~UT). 
From the estimations of the height of the EUV wave crest and its orthogonal projections, we determine the velocity over the period 5:45--5:55~UT, and the results are listed in Table~\ref{Table6}. The velocity of the EUV wave crest determined  on the basis of line-of-sight projection is about 25\% larger (259--261 km/s) than that using the orthogonal projection (205--208 km/s), indicating that in this case the EUV wave velocity determined on the basis of the line-of-sight projection is overestimated. This is related to the increase of the EUV wave front height from 54 to 93~Mm during the considered interval 5:45--5:55~UT, and thus the line-of-sight projections propagated over a larger distance than the orthogonal ones.
\begin{deluxetable}{c c c c c}
	\tablecaption{Estimations of EUV wave kinematics on 13 February 2009 5:45--5:55~UT.
		``Time'' gives the time period over which the velocity of EUV wave is estimated.
		``Point'' indicates a number of points $N$ and $M$ (brown points, Figure~\ref{fig10}). ``Velocity (km/s), Orthogonal projection'' gives the estimation of velocity using the orthogonal projections of the wave crest onto the surface. ``Velocity (km/s), Line-of-sight projection'' shows the estimation of velocity
		on the basis of line-of-sight projection.  
		\label{Table6} } 
	\tablehead{
		\colhead{Time} &
		\colhead{Points} &
		\colhead{\parbox{4cm}{\centering Velocity (km/s) \\ Orthogonal projection}} &
		\colhead{\parbox{4cm}{\centering Velocity (km/s) \\ Line-of-sight projection}} &
		\colhead{Sector} 
	}
	\startdata
	5:45 - 5:55~UT    & 1  & 208 & 261 & 19\\ 
				 	  & 2  & 208 & 259 & 19\\
                   	  & 3  & 205 & 259 & 18\\ 
	\enddata                 
\end{deluxetable}

\section{Discussion and Conclusions} 
In this paper we have developed techniques to reconstruct the 3D structure of EUV wave fronts using 2D projected multi-point observations from the STEREO/EUVI-A and EUVI-B instruments using the 195\AA~filter, where the contrast of the transient is highest. The 3D reconstruction of the heights of the EUV wave fronts allows us also to correct the wave kinematics for projection effects. These techniques are applied to the study of two events:
on 7 December 2007 (AR 10977), where the  STEREO spacecraft separation was 42$^\circ$, and on 13 February 2009 (AR 11012), where the two STEREO spacecraft were in quadrature. 
 
Epipolar geometry and triangulation are very useful methods for 3D reconstruction. However, they do not give an automatic solution for the matching problem. We have developed two different methods to solve the matching problem, i.e.\ to find the correspondence between points on the EUV wave crest in the STEREO-pair images. The two methods are suitabe for the different stages of EUV wave development, i.e.\  for the early evolutionary phase where we observe distinct wave fronts and for the later stages where the EUV wave fronts have become diffuse and broadened. 
At the early stage of development when two STEREO spacecraft observe distinctly different facets of the wave and associated dimming regions, we used the techniques of epipolar geometry of stereo vision in combination with the analysis of intensity perturbation profiles. This combination allows us to automatically determine the location of the crest of the EUV wave front and to estimate its height.  We also demonstrated that there are wave front segments in which the correspondence problem becomes ill-conditioned and degenerates, and thus the calculation of wave heights in these regions would lead to very uncertain results.  Note that ill-conditioned zones are a general problem of epipolar geometry and 3D reconstructions,
and careful consideration of the sectors in which the 3D wave structure and propagation can be reliably reconstructed is needed. Therefore, we applied our analysis only to selected segments of the wave crest on both images, in which the problem is well-conditioned.

To derive the heights of the EUV wave crest during the later stages of the event, a ring analysis technique was proposed. At the later stages of evolution, the wave fronts become more diffusive and faint. However, they also become more similar in their appearance as observed by the two STEREO spacecraft, and this similarity can be used to solve the matching problem.  In the ring analysis, we calculate the integrated pixel intensities of small regions belonging to the wave crest on both the STEREO-A and STEREO-B images, to automatize the process of finding the correspondence between related points.  

In our study, we focused on segments of the wave front, which were undisturbed by the interaction with other plasma structures like ARs or coronal holes, and which were unaffected by ill-conditioning. For the event on 7 December 2007, the reconstructed  height of the wave front at 4:35~UT is of 90--104~Mm, and decreases to 7--35~Mm in the decay phase during 4:45--4:55~UT. At 4:35~UT, the orthogonal projections of the wave crest onto the surface are on average 20\% closer to the eruptive center than the line-of-sight projections. Later at 4:45--4:55~UT, the difference between orthogonal and line-of-sight projections becomes insignificant because the height of the EUV wave front has strongly decreased. As the dynamics of orthogonal projections determines the velocity of an EUV wave on the sphere, these findings should be taken into account for the calculations of EUV wave kinematics. We calculated the corrected EUV wave kinematics  on the basis of the orthogonal projections of the wave crest onto the solar surface. We found that during 4:35--4:55~UT, the velocity of the EUV wave crest on the sphere changes from 217 to 266 km/s. During this period, the velocity of the wave crest estimated on the basis of line-of-sight projection is underestimated by 5--18\%. This is related to the fact that, with increasing wave front height the distance between orthogonal and line-of-sight projections increases. With the decay of the wave front height to 7--35~Mm at 4:45--4:55~UT, the line-of-sight projections became close to the orthogonal ones, and the velocities derived from the orthogonal and line-of-sight projections are similar. 

For the 13 February 2009 event, we find that the height of the EUV wave crest increased from 54 to 93~Mm over 10 minutes from 5:45 to 5:55~UT, and the EUV wave velocity calcuated on the basis of the dynamics of the orthogonal projections is 205--208 km/s. In this event, we find that the velocity of the wave crest as derived from line-of-sight projection is overestimated by $\sim$25\%. This difference is related to the strongly increasing EUV wave front height during the considered time range, and thus the distance between the orthogonal and line-of-sight projections also increased. These findings suggest that the 3D reconstructions of EUV wave front heights are an important factor, in order to derive reliable estimates of the EUV wave front speed and its changes during the wave evolution. Depending on the evolution of the EUV front height in the time range considered, i.e.\ either increasing or decreasing, ignoring this effect will lead to and over- or underestimations of its propagation speed, and may thus result in non-physical variations in the derived evolution of the EUV wave speed in terms of acceleration/deceleration. 

\citet{Kienreich2009} have also estimated the EUV wave height for the 13 February 2009 quadrature event. Their analysis is based on comparing the EUV kinematics as derived on the limb and on the disk by the two STEREO spacecraft, resulting in a height in the range of  80--100~Mm. This estimate is consistent with our finding for the wave height obtained at 5:55 UT. 
\citet{patsourakos09} studied the EUV wave front height for the 7 December 2007 event when the STEREO spacecraft separation was 42$^\circ$. They applied a triangulation to segments of the EUV wave front, visually identified by point-and-click  at 
4:45~UT, when the observations from both spacecrafts were already similar, and obtained a height of about 90~Mm. This is different to the wave front height we obtain at this time, where according to our analysis the wave front height has already decayed from 90--103 Mm at 04:35 UT  to 7--35~Mm at 4:45 UT.  We note that segments of the wave front analysed in \citet[][see their Figure 8]{patsourakos09} lies in the ill-conditioned zone (the segments in the direction of NorthWest), where the obtained solution is not unique. 
The only other study we are aware of, that derived the height of EUV wave fronts not only statically but also how it changes over the event evolution, is \cite{Delannee2014}. These authors performed a comprehensive and in-depth analysis of the 7 December 2007 event, using three different reconstruction methods (triangulation, reverse projection differencing, reverse projection correlation). They find that the wave front height is initially increasing during the first 5 min (from 34 to 74 Mm), and then decays back to the low corona. This is consistent with our findings of the evolution of the wave front height in this event. 

In summary, the present and previous studies \citep{Kienreich2009,Patsourakos2009b,Delannee2014} suggest that the heights of the EUV wave front in its main phase are mostly in the range of 70--100 Mm. This is in the range of the coronal scale height for plasma temperatures of about 1.5--2.0 MK. As discussed in \cite{Patsourakos2009b}, this  is consistent with the propagation of a fast-mode wave over quiet Sun regions, as it is expected that  the passing wave front perturbs the ambient coronal plasma with the bulk of the plasma confined within a coronal scale height. However, as shown in the present study as well as in \cite{Delannee2014}, during the decay phase of the EUV waves the reconstructed heights are considerably lower. 
This may be related to the decay of the perturbation profile amplitude during the EUV wave evolution \citep{veronig2010,warmuth15,Vrsnak2016}. 
Finally, we note that the EUV waves to which these STEREO 3D reconstructions were applied, were all rather weak and slow events that occurred during solar minimum conditions. Observations of very strong EUV waves observed above the solar limb, reveal that the EUV wave fronts may be observable up to $>$0.5 solar radii \citep{Kwon2013,Liu2018,Veronig2018}, though also in these cases the bulk of the emission is observed at low coronal heights.

\begin{acknowledgements}
A.M.V., K.D. and M.T. acknowledge the Austrian Science Fund (FWF):
P24092-N16 and the Austrian Space Applications Program of the Austrian 
Research Promotion Agency FFG (ASAP-11 4900217).
The STEREO/SECCHI data are produced by an international consortium of the Naval Research Laboratory (USA), Lockheed Martin Solar
and Astrophysics Lab (USA), NASA Goddard Space
Flight Center (USA), Rutherford Appleton Laboratory
(UK), University of Birmingham (UK), Max-Planck-Institut f{\"u}r Sonnenforschung (Germany), Centre Spatiale de Li{\`e}ge (Belgium),
Institut d'Optique Th{\'e}orique et Appliqu{\'e}e (France),
and Institut d'Astrophysique Spatiale (France).
\end{acknowledgements}


\end{document}